\documentclass{styles/svproc}
\usepackage{url}

\usepackage{amsmath}
\usepackage{graphicx}
\usepackage{enumitem}
\usepackage{url}
\usepackage{tipa}

\begin{document}
\mainmatter              
\title{Explaining Spectrograms in Machine Learning: A Study on Neural Networks for Speech Classification}
\titlerunning{Explaining Spectrograms in Machine Learning}  
%

\author{Jesin James\inst{1} \and Balamurali B. T.\inst{2} \and
Binu Abeysinghe\inst{1} \and Junchen Liu\inst{1}}

\authorrunning{Jesin James et al.} 
%
\tocauthor{Jesin James\, Balamurali B. T., Binu Abeysinghe, Junchen Liu}
\institute{Department of Electrical, Computer, and Software Engineering, \\ The University of Auckland, New Zealand\\
\email{jesin.james@auckland.ac.nz},\\
\and
Singapore University of Technology and Design, Singapore}

\maketitle              

\begin{abstract}
This study investigates discriminative patterns learned by neural networks for accurate speech classification, with a specific focus on vowel classification tasks. By examining the activations and features of neural networks for vowel classification, we gain insights into what the networks ``see" in spectrograms. Through the use of class activation mapping, we identify the frequencies that contribute to vowel classification and compare these findings with linguistic knowledge. Experiments on a American English dataset of vowels showcases the explainability of  neural networks and provides valuable insights into the causes of misclassifications and their characteristics when differentiating them from unvoiced speech. This study not only enhances our understanding of the underlying acoustic cues in vowel classification but also offers opportunities for improving speech recognition by bridging the gap between abstract representations in neural networks and established linguistic knowledge.
\keywords{Spectrograms, Linguistics, Explainable Speech Recognition, Interprettable, Activation Maps, Vowels }
\end{abstract}
\section{Introduction}
\label{sec:intro}

In recent years, the field of speech recognition has experienced remarkable progress, primarily driven by the widespread adoption of deep neural networks (DNNs) to train speech recognition models. The successful application of DNNs in speech recognition has led to significant advancements in various domains such as automatic speech recognition, voice assistants, and language understanding. Spectrograms, which provide a visual representation of the frequency content of a speech signal as it evolves over time, offer a promising alternative to conventional speech representations in the context of DNN-based speech recognition. Convolutional neural network (CNN) is a type of DNNs originally designed for image processing. However, CNNs have been successfully adapted to process spectrograms, capturing temporal dependencies and extracting meaningful features. Also, the field of computer vision using neural networks has progressed extensively with large networks such as ResNet \cite{resnet}, VGGNet \cite{simonyan2014vggnet},  DenseNet \cite{huang2017densenet} and similar trained on large image datasets. Some of these large networks have been used for speech recognition tasks with spectrograms as the input. For example, ResNet, with its ability to handle deep architectures and alleviate the vanishing gradient problem, has also been explored for speech tasks such as speech recognition \cite{tang2018ResNetexample,zou2020end_resnetexample}, keyword spotting \cite{tang2018deep_ResNetkeyword} and emotion recognition \cite{triantafyllopoulos2019_renet_emotion}. 

The choice of using spectrograms in speech recognition holds several advantages. Spectrograms effectively capture both the temporal and spectral information contained in speech, offering a comprehensive representation that enables the model to discern important acoustic cues for various application. Additionally, spectrograms provide a visual understanding of the speech signal, allowing researchers to interpret and analyze the underlying speech patterns more intuitively. However, despite the benefits, one critical challenge arises: the lack of understanding regarding what the model is learning from spectrograms. While humans can manually annotate and recognize speech components from spectrograms, the extent to which a model is learning the same features is not always clear. This poses the following issues for current and future research on speech recognition tasks:

\textbf{Interpretability and Performance Improvements:} Despite the impressive results achieved by these models, they are often regarded as black boxes, which limits researchers and practitioners from gaining a deep understanding of the specific features and patterns that contribute to their decision-making process. This lack of transparency and interpretability hinders further improvements in model performance.


\textbf{Unoptimised Model Training:} 
The process of human annotation of spectrograms relies heavily on linguistic insights, allowing annotators to focus on specific aspects relevant to speech analysis. However, many DNNs used in speech recognition are not optimized using linguistics knowledge, leading to unoptimised model training. These neural networks often treat spectrograms as mere `images', lacking a comprehensive understanding of the frequency axis and its significance in speech analysis. As a result, the model's ability to refine and optimize speech recognition is limited.


This study addresses the above knowledge gaps by investigating what neural networks learn from spectrograms. Focusing on two specific problems, namely vowel classification and voiced-unvoiced classification, we aim to unravel the black box nature of neural networks trained on spectrograms. The main contributions of this paper are:


\begin{enumerate}[leftmargin=*]
\vspace{-0.3cm} 
\item \textit{Designing experiments} to explore the relationship between neural networks' learning from spectrograms and human interpretation of spectrograms.

\item \textit{Employing visualising techniques} to identify  the regions of a spectrogram that are considered `important' by DNNs in specific speech recognition tasks.

\item \textit{Explaining} the results to uncover insights into DNN's understanding of spectrograms in relation to human interpretation of the same.

\end{enumerate}

In this study, \textit{interpretation}, refers mapping an abstract concept, like a predicted class, into a human-understandable form such as images or texts. An \textit{explanation} consists of interpretable features that contributed to a classification or regression decision. An example is a heatmap highlighting the pixels in an image that strongly support the decision \cite{montavon2018methods_interpreting}.  


\section{Background and Related Work}
\begin{figure*}[t]
	 \vspace{-0.4cm}
		\centering
		\centerline{\includegraphics[scale=0.5]{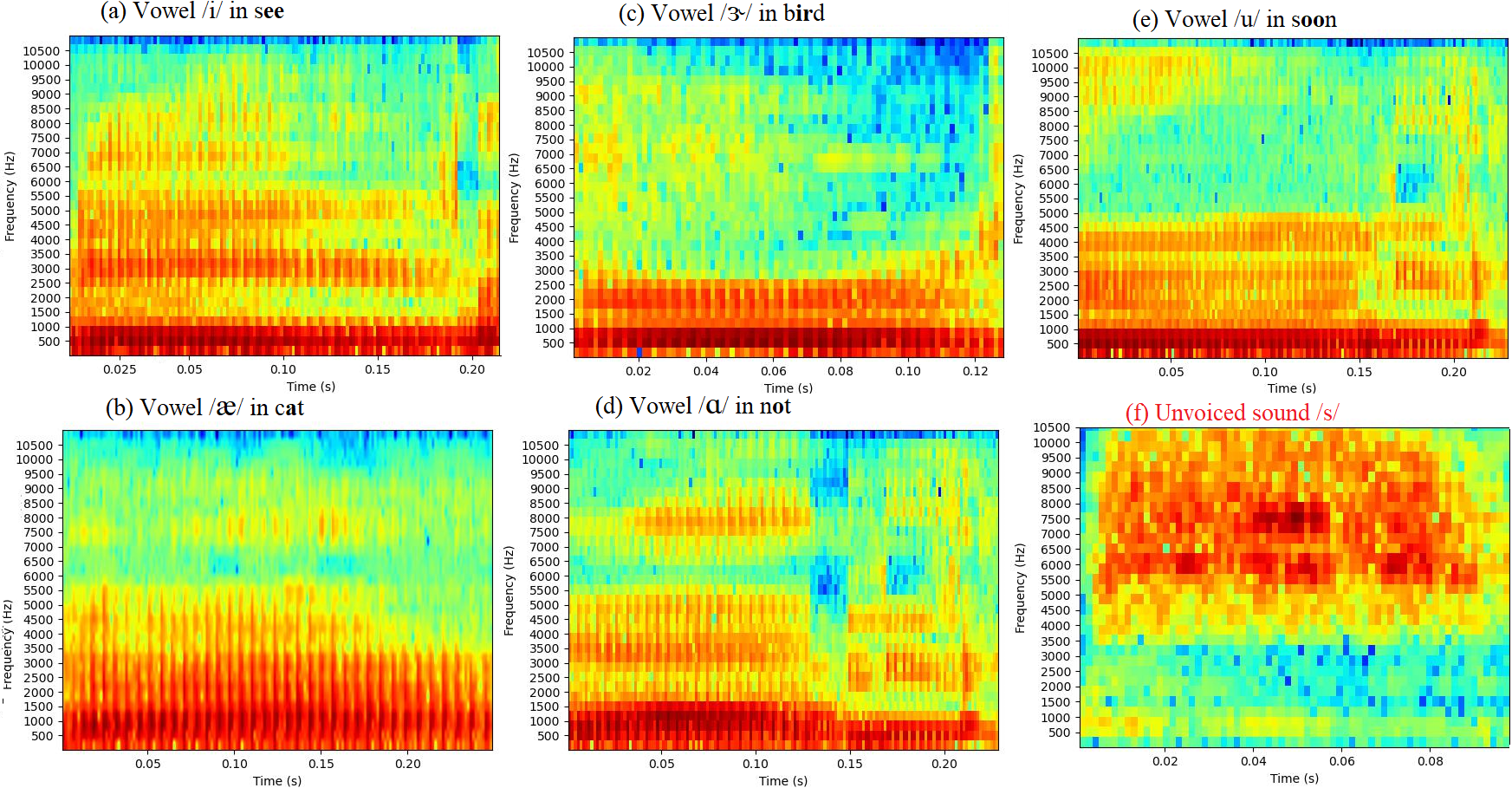}}
   \vspace{-0.4cm}
		\caption{Spectrograms of five American English Vowels, which are voiced sounds and unvoiced sound /s/.}
	\label{fig:vowels_spectrogram}
 \vspace{-0.4cm}
\end{figure*}
\subsection{Spectrograms and their Significance in Linguistics}
\label{sec:spectrogram_lingusitics}

\vspace{-0.2cm}

Spectrograms provide visual representations of speech signal frequencies over time. They are crucial in linguistics for analyzing acoustic properties of speech and offer detailed insights into temporal and spectral characteristics, allowing researchers to study articulatory gestures and acoustic cues.


In phonetics, spectrograms are instrumental in studying speech and its production \cite{flanagan2013speech_spectrogram}. They provide a tool to analyze the fine-grained details of articulatory movements, such as formant patterns and consonant releases \cite{spectrogram_reading2}. Spectrograms allow linguists to investigate phonetic features like voicing, place and manner of articulation, and vowel quality, helping in the characterization and classification of speech sounds across languages \cite{spectrogram_reading,spectrogram_reading1}. There are even spectrogram reading competitions in conferences such as International Phonetics Association Conference and Australasian Speech Science and Technology Association Conference. 

In phonology, spectrograms aid in understanding phonological processes and patterns \cite{ladefoged_spectrogram_phonology}. They facilitate the identification of phonemic contrasts, allophonic variations, and sociolinguistic phenomena such as regional accents and dialectal variations \cite{holmes2013research_spectrogram}. Spectrograms also aid in the study of language variation, speech disorders, language acquisition, and cross-linguistic differences in phonetic patterns. While spectrograms have limitations in capturing prosody, intonation, and discourse structure, they remain highly significant in linguistics as a powerful tool for analyzing speech sounds.


\vspace{-0.4cm}
\subsubsection{Voiced and Unvoiced Speech}

Voiced speech is produced when the vocal folds vibrate, resulting in a periodic airflow. They include sounds such as vowels and voiced consonants. In spectrograms, voiced sounds exhibit a characteristic pattern as seen in Fig. \ref{fig:vowels_spectrogram}(a) to (e). They display regular bands of energy, known as formants, which represent the resonant frequencies of the vocal tract. Voiced speech has a harmonic structure and exhibit sustained energy throughout their duration. 

Unvoiced sounds are produced without vocal fold vibration. They include voiceless consonants. The spectrograms of unvoiced sounds lack a clear harmonic structure and exhibit a more random and dispersed distribution of energy across a wide range of frequencies, as seen in Fig. \ref{fig:vowels_spectrogram}(f). Unvoiced sounds are characterized by transient bursts of energy concentrated around the onset and release of the sound \cite{Russel_online_spectrograms}. 

\vspace{-0.4cm}
\subsubsection{Vowel Sounds}

The combination of formant information and supplementary spectral features in spectrograms enables linguists to distinguish and analyze vowel sounds in their linguistic investigations. Formants correspond to the resonant frequencies of the vocal tract during vowel production. By observing the positioning, spacing, and relative intensity of these formants in the spectrogram, linguists can identify and categorize different vowel sounds. For example, in Fig. \ref{fig:vowels_spectrogram} (a) to (e), the vowels have distinct formant patterns. The first formant of /\textipa{i}/ is 385 Hz, /\textipa{u}/ is 400 Hz, /\textipa{\ae}/ is 800 Hz, /\textipa{\textrhookrevepsilon}/ in 590 Hz and  /\textipa{A}/ is 710 Hz (formant estimation was done by observing the spectrograms and verified using Praat \cite{boersma2001praat}).  These values are within the frequency range expected for American English \cite{rabiner1978digital_formants}. Differences exist for the second, third and fourth formants too, as seen by the formant bands at different frequencies in the Fig.. 

Spectrograms offer insights into acoustic cues related to vowel articulation, including vowel duration and spectral shape. The shape of the spectral pattern in a vowel is influenced by the tongue position and the openness of the oral cavity, which determine the shape and configuration of the vocal tract. High vowels like /\textipa{i}/ and /\textipa{u}/ typically exhibit a more concentrated spectral shape with higher energy in the higher frequency range (See \ref{fig:vowels_spectrogram} (a) and (e)). This results from the tongue being positioned closer to the roof of the mouth, giving rise to a narrower constriction in the vocal tract and emphasizing higher-frequency resonances. The distinction in vowel duration can be observed using the spectrogram's horizontal axis. E.g., comparing the horizontal axis of Fig. \ref{fig:vowels_spectrogram} (a) and (c), we can observe /\textipa{i}/ is a longer vowel than /\textipa{\textrhookrevepsilon}/.


\vspace{-0.4cm}
\subsection{Explaining what Deep Neural Networks Learn}

Researchers have employed various methods to gain insights into the learning process of deep neural networks \cite{montavon2018methods_interpreting}. Visualization techniques, such as highlighting the important areas in an image for a specific prediction, are commonly used \cite{simonyan2013deep_visual,landecker2013interpreting_visual}. Other approaches include sensitivity analysis, Taylor decomposition, and backward propagation techniques \cite{montavon2018methods_interpreting}.


Class activation maps (CAMs) are visualization techniques to explain the decision-making process of deep neural networks, specifically in CNNs for image classification tasks \cite{zhou2016learning_CAM,simonyan2013deep_visual,landecker2013interpreting_visual}. CAMs provide valuable insights into the influential regions within an image that contribute to predicting a particular class label. By leveraging the gradients during the backward propagation process, CAMs capture the importance of spatial locations, highlighting the regions that significantly influence the final classification decision. These maps have proven to be effective in explaining the reasoning behind deep learning models, allowing researchers and practitioners to comprehend which areas of an image play a crucial role in making accurate predictions. CAMs offer a visual explanation by generating heatmaps that emphasize the relevant regions responsible for the classification decision.


Explanations of deep learning models have been widely explored in various domains, including image classification, pattern recognition \cite{bai2021explainable_ptterm}, and medical applications \cite{holzinger2019causability_medicine,roy2020deep_medicine}. While activation mapping has been extensively applied in visual recognition tasks, its direct application to speech recognition is less common. One example is the use of CAM to explain the results of detecting oral cancer speech using a ResNet-based classifier with spectrograms as input \cite{halpern2020detectingcancer}.

In this study, we propose adapting the concept of CAMs to spectrograms with the aim of identifying the specific frequency regions that provide the most informative cues for speech classification tasks.


\vspace{-0.4 cm}

\section{Methodology}
\label{sec:format}



\vspace{-0.4 cm}

\subsection{Database}

The LJSpeech corpus (\cite{ljspeech17}) was selected as the database for training and evaluating the classification models in this study. This corpus consists of American English recordings by a speaker who identifies as female, along with text transcriptions, all sampled at 22,050 Hz. To align the recordings with their respective transcriptions WebMAUS \cite{kisler2017multilingual_webmaus} was utilized with American English option. 

We limit the scope of the study to five vowels chosen to span over the American English vowel space \cite{americanenglish}: /\textipa{i}/  (high, front),  /\textipa{\ae}/ (low, front), /\textipa{\textrhookrevepsilon}/ (mid), /\textipa{A}/ (low, back) and  /\textipa{u}/ (high, back). For each vowel, start and end times were identified, and the appropriate segments were extracted. The resulting dataset comprised a total of 79,269 single-vowel recordings.


Unvoiced consonants were also extracted from LJSpeech corpus. The selected consonants were /\textipa{p}/, /\textipa{t}/, /\textipa{k}/, /\textipa{f}/, /\textipa{s}/, /\textipa{t\textesh}/, /\textipa{\textesh}/, /\textipa{$\theta$}/. Only 79,269 instances of these consonants were included to match the number of vowels.

\newpage
\subsection{Experiment Design}

The methodology encompasses three experiments: 
\begin{enumerate}[leftmargin=*]
\vspace{-0.2cm} 
\item \textbf{Vowel classification} using all frequency components present in the speech signal

\item \textbf{Vowel classification }focusing on the region containing formants, i.e. 4000 Hz

\item \textbf{Voiced vs unvoiced classification}  focusing on the region containing formants, i.e. 4000 Hz
\vspace{-0.2cm}
\end{enumerate}

The first experiment assesses the classification model's ability to identify relevant patterns in the spectrogram using speech in LJSpeech corpus considering all frequency components present, i.e. upto sampling frequency/2 =  11,025 Hz. Due to linguistics knowledge that the first four formants of the selected vowels have frequency less than 4000 Hz, the second experiment restricts the maximum frequency to 4000 Hz. This limitation narrows down the scope of visual representation provided to the network, allowing for a more focused analysis of vowel characteristics already used by linguists. The final experiment investigates the network's capability to accurately distinguish between voiced and unvoiced sounds. From linguistic knowledge, we know that the existence of fundamental frequency is a distinguishing feature between these two categories \cite{rose2002forensic}. This experiment aims to examine whether the model would accurately identify fundamental frequency  along with other relevant frequency of importance in differentiating between voiced and unvoiced sounds.

\subsection{Classification Model Training}

For this study, the ResNet-101 model was used. ResNet-101 is an enhanced version of the original ResNet \cite{resnet} with 101-layers, addressing issues related to network depth and degradation by employing residual learning frameworks. ResNet-101 already pretrained on the ImageNet dataset \cite{imagenet_cvpr09} was subsequently fine-tuned using the vowel and consonant datasets mentioned earlier. Three instances of ResNet-101 were fine-tuned for this purpose, one for each experiment employing a 70/30 train-test split, each using softmax activation function to make the final decision. The hyperparameters remained consistent with the base ResNet-101 model, with a batch size of 32, learning rate of 0.0001, and Adam as the optimizer. The training process was conducted on a local machine equipped with two Nvidia RTX 3090 GPUs, each possessing 24GB of VRAM. Training time amounted to approximately 3.5 hours each for Experiment 1, 2 and 6 hours for Experiment 3.

\subsection{Class Activation mapping (CAMs)}
\begin{figure*}[p]
        \centering
        \includegraphics[scale = 0.2]{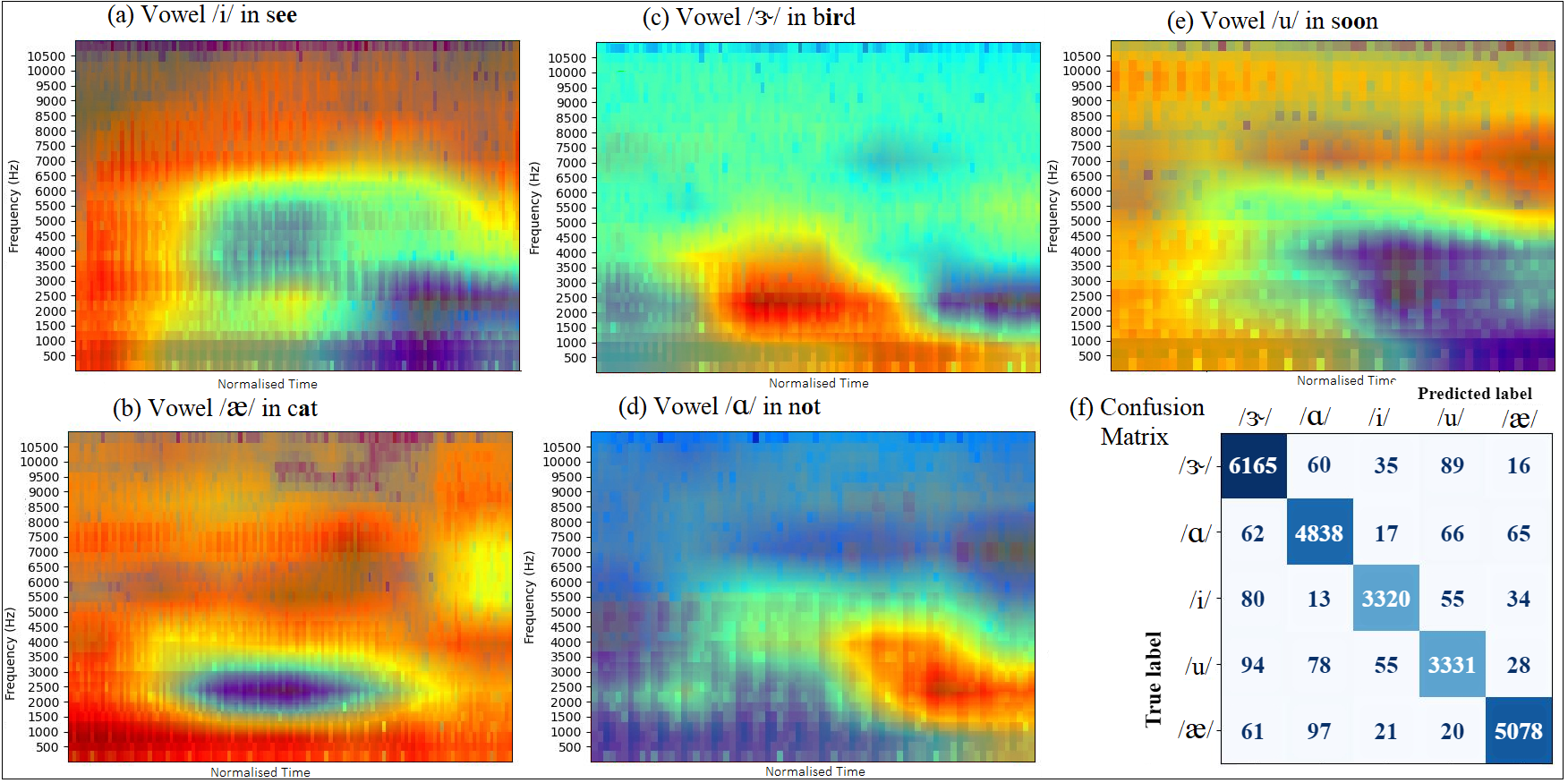}
        \vspace{-0.4cm}
        \caption{CAMs and Confusion Matrix for classifying five vowels with maximum frequency (Sampled at 22050 Hz)).}
	\label{fig:vowels_CAM_2205k}
 
      \vspace{0.1cm}
        \includegraphics[scale = 0.2]{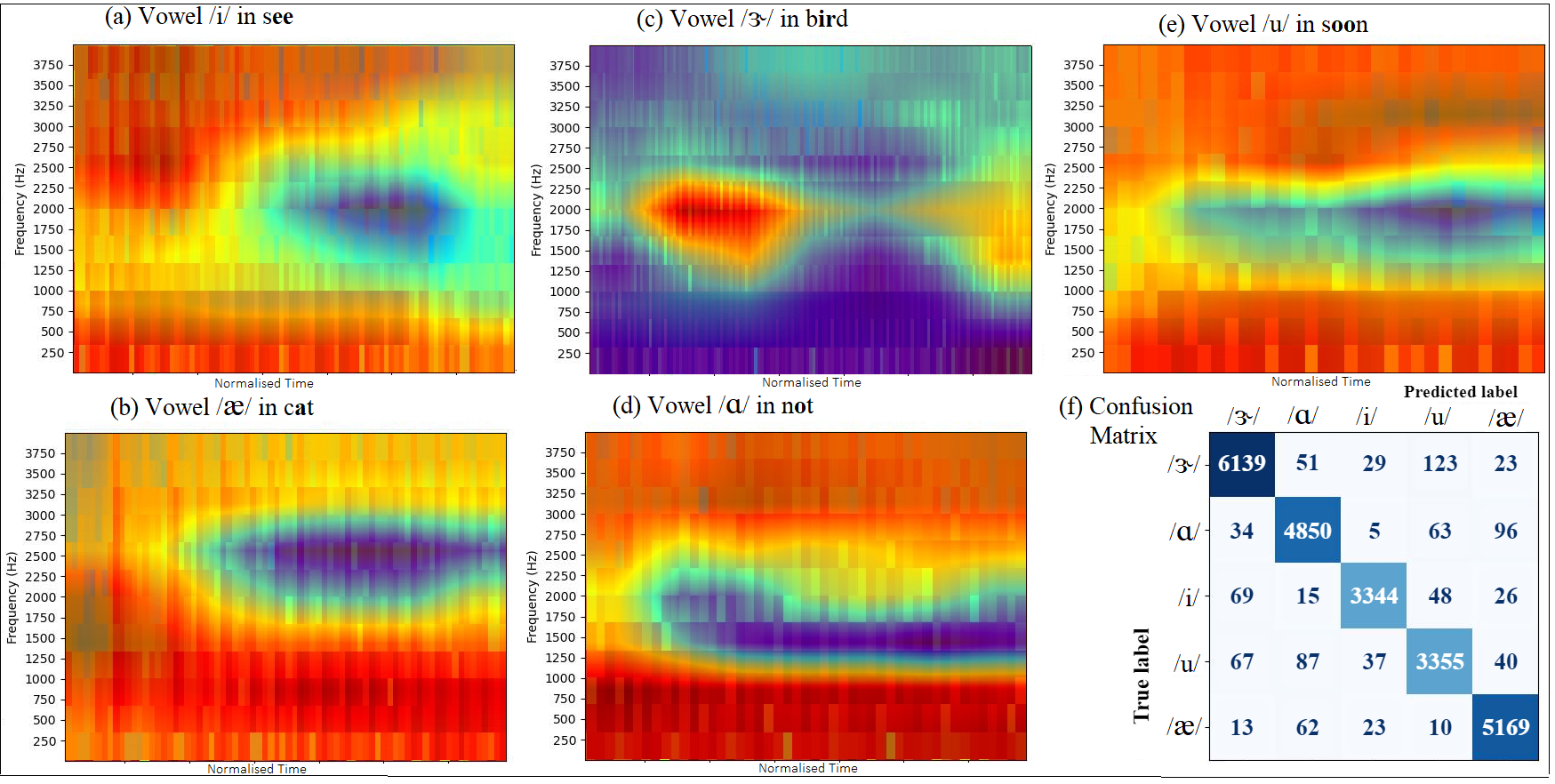}
       \vspace{-0.4cm}
        \caption{CAMs and Confusion Matrix for classifying five vowels with frequency limited to 4000 Hz.}
	\label{fig:vowels_CAM_4k}

        \includegraphics[scale = 0.2]{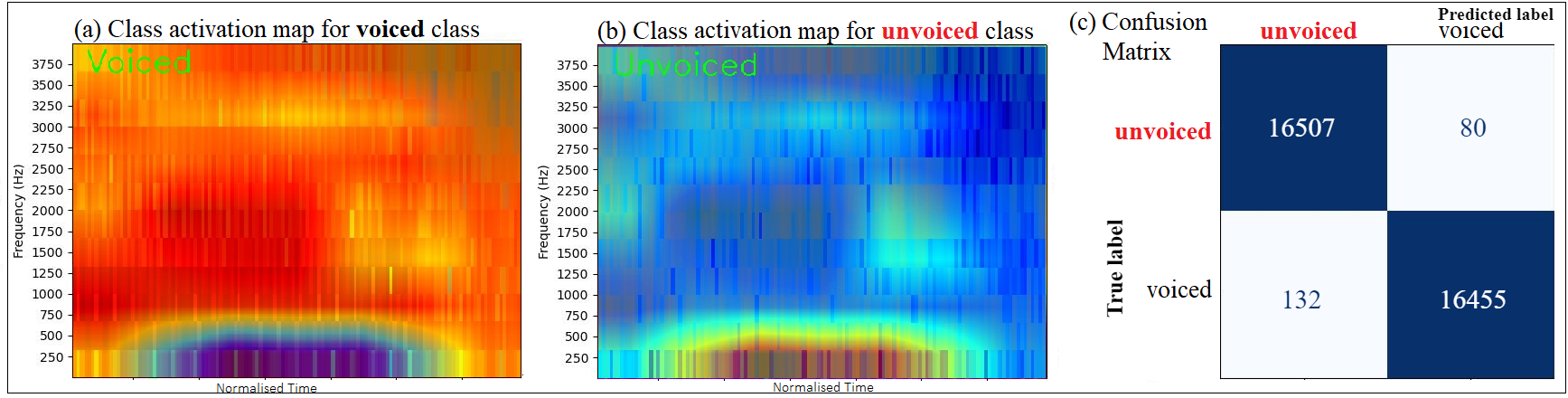}
        \vspace{-0.5cm}
        \caption{CAMs and Confusion Matrix for classifying voiced and unvoiced speech with maximum frequency =  4000 Hz.}
	\label{fig:vowels_CAM_unvoiced}
    \end{figure*}

The three models were compared using CAMs, using the approach reported in \cite{zhou2016learning_CAM}. CAMs provide insight into how the model's focus shifts when presented with different inputs while keeping the model architecture consistent. To generate the CAMs, first the spectrograms for each class were resized to 224 $\times$ 224. Then, the dot product between the final convolutional layer's feature for a class and the softmax weights of the output layer was calculated. The result was normalized and scaled to the same resolution. Finally, the resulting array was transformed into a heatmap. The heatmap was then superimposed on the spectrogram of a speech signal to obtain the CAMs in Figures \ref{fig:vowels_CAM_2205k}, \ref{fig:vowels_CAM_4k}, \ref{fig:vowels_CAM_unvoiced}. The code for training ResNet-101 and generating CAMs is made available\footnote{\url{https://github.com/MaoriEnglish-Codeswitch/Vowel_Classification}}.


\vspace{-0.4 cm}
\section{Results and Discussion}
\label{sec:results}

\subsection{Vowel classification using all Frequency Components}
\subsubsection{Class Activation Map Analysis}
The CAM analysis provides insights into the discriminative properties of the considered vowels as seen in Fig. \ref{fig:vowels_CAM_2205k}). Darker red regions in the CAM, indicating higher importance, were predominantly observed in the high-frequency region for three vowels: frequency $>$ 5500 Hz for /\textipa{i} and /\textipa{u}/, but $>$  3500 Hz for /\textipa{\ae}/ (Fig. \ref{fig:vowels_CAM_2205k} (a), (b), (e)). Among these, the vowel /\textipa{i}/ is observed to be the darkest in the high-frequency region, followed by /\textipa{\ae}/ and /\textipa{u}/. These observations suggest that high-frequency components play a crucial role in distinguishing these vowels. The presence of high energy in the high-frequency region of their spectrograms in Fig. \ref{fig:vowels_spectrogram} (a), (e) likely contributes to their distinctiveness in the CAMs. This is also expected as both /\textipa{i}/ and /\textipa{u}/ are high vowels having high energy in high frequency range.

Interestingly, for the /\textipa{\ae}/ vowel in Fig. \ref{fig:vowels_CAM_2205k} (b), an additional region of importance was identified in the low-frequency range. Specifically, a strong band of frequencies between 500 and 1000 Hz exhibited a darker region in the CAM. This finding indicates the potential role of this frequency band in predicting the presence of the /\textipa{\ae}/ vowel and this region corresponds to the first formant of /\textipa{\ae}/, as seen in Fig. \ref{fig:vowels_spectrogram} (b) .

In contrast, the  /\textipa{\textrhookrevepsilon}/and /\textipa{A}/ displayed similar characteristics in the high-frequency region (frequency $>$ 5000 Hz) as in Fig. \ref{fig:vowels_CAM_2205k} (c) and (d), with no prominent dark red heatmap. However, in the low-frequency range of 500 to 1000 Hz, as seen in Fig. \ref{fig:vowels_CAM_2205k} (c) the vowel /\textipa{\textrhookrevepsilon}/ demonstrated a darker region that appeared in the latter half of the spectrogram. On the other hand, the vowel /\textipa{A}/ exhibited a similar trend, albeit with the frequency of interest slightly higher, ranging from 1000 to 3500 Hz.

\vspace{-0.4 cm}
\subsubsection{Confusion Matrix Analysis}
Analysis of the confusion matrix revealed a high overall accuracy $>$ 96\% as seen in the resulting confusion matrix  in Fig. \ref{fig:vowels_CAM_2205k} (f). However, some minor misclassifications were observed. The vowel /\textipa{A}/ misclassified as /\textipa{i}/ was found to be the lowest, which can be attributed to the disjoint nature of their CAMs. The distinct patterns in the CAMs for /\textipa{A}/ and /\textipa{i}/ contribute to their accurate discrimination seen in Fig. \ref{fig:vowels_CAM_2205k} (a) and (d).

Misclassifications were observed among the vowels /\textipa{\textrhookrevepsilon}/, /\textipa{A}/, and /\textipa{u}/. Some of /\textipa{\textrhookrevepsilon}/ were misclassified as /\textipa{A}/ and /\textipa{u}/, and similar misclassification were observed in /\textipa{A}/ v.s. /\textipa{\textrhookrevepsilon}/ and /\textipa{u}/ v.s. /\textipa{\textrhookrevepsilon}/. This similarity in misclassifications can be attributed to the shared characteristics observed in the low-frequency region of the CAMs for /\textipa{\textrhookrevepsilon}/, /\textipa{A}/, and /\textipa{u}/ as seen Fig. \ref{fig:vowels_CAM_2205k} c, d and e, respectively. Although /\textipa{\textrhookrevepsilon}/and /\textipa{u}/ appear almost disjoint in their activation maps, a small overlap in the region of importance in the low frequency range of 500 to 1000 Hz, around the lower-middle part as seen in Fig. \ref{fig:vowels_CAM_2205k} (c), (e) could potentially result in the misclassifications.

Similar misclassifications were also observed between /\textipa{A}/ and /\textipa{u}/ , as well as /\textipa{A}/ and /\textipa{\ae}/. However, these misclassifications cannot be easily explained using the CAMs of /\textipa{A}/ vs /\textipa{u}/ or /\textipa{A}/ vs /\textipa{\ae}/. However, there is a small region of overlap in the latter case, specifically in the 1000 to 3500 Hz region towards the latter part of the CAM, which could have partially accounted for the misclassifications.

A number of Vowel /\textipa{i}/ was misclassified as /\textipa{\textrhookrevepsilon}/ and /\textipa{u}/. Surprisingly, this misclassification response was found to be non-symmetric for /\textipa{\textrhookrevepsilon}, i.e., the number of /\textipa{\textrhookrevepsilon}/ misclassified as /\textipa{i}/ is marginally less however for the case of /\textipa{A}/ and /\textipa{u}/, there are still misclassification and the misclassification is symmetric. The misclassification between /\textipa{i}/ and /\textipa{\textrhookrevepsilon}/ cannot be explained using the CAMs alone. However, the similarity between the activation maps of /\textipa{i}/ and /\textipa{u}/ in the high frequency region could contribute to the higher number of misclassifications. The non-symmetric misclassification (for e.g. observed between /\textipa{\textrhookrevepsilon}/ and /\textipa{\ae}/), could be attributed to mislabelling or possibly noisy speech data needing further investigation.

\vspace{-0.4 cm}
\subsection{Vowel Classification focusing on Region of Formants}
\label{sec:expt2}

As seen in Fig. \ref {fig:vowels_CAM_4k}, the comparison between CAMs obtained by focusing on the region upto 4000 Hz and ones with all frequencies revealed some differences in the importance regions.

For /\textipa{\ae}/, the CAM remained similar between the two.

For vowel /\textipa{\textrhookrevepsilon}/, a similarity was observed in the CAM's shape. But the importance seen in the low frequency region for when considering all frequencies shifted upward. The red region shifted from 500 Hz - 1000 Hz in Fig. \ref{fig:vowels_CAM_2205k} (c) to 1500 Hz - 2500 Hz in Fig. \ref{fig:vowels_CAM_4k} (c). On the other hand, for the vowel /\textipa{u}/ and /\textipa{i}/, a new region of importance emerged in the low frequency region around 500 to 1000 Hz.

The CAM for the vowel /\textipa{A}/ exhibited significant differences between the two cases. The region of importance appeared completely swapped between the original sampling frequency case and the frequency limited cases.

Despite the subtle differences observed in the CAM between the original and frequency limiting cases, the resulting confusion matrices showed remarkable similarity. Misclassifications of /\textipa{A}/ as /\textipa{i}/ still resulted in the lowest number, possibly due to the presence of a region of high importance in /\textipa{A}/ as a strong band observed around frequencies $<$ 1000 Hz in /\textipa{A}/ but not in /\textipa{i}/). However,  similarities were observed in the high-frequency region of their CAMs.

The vowel /\textipa{\ae}/ and /\textipa{A}/ showed similar CAMs, and it was not surprising to find a slightly higher misclassification rate between /\textipa{\ae}/ and /\textipa{A}/ and vice versa in this investigation. Interestingly, /\textipa{\textrhookrevepsilon}/ and /\textipa{u}/ exhibited disjoint CAM, with minimal overlap in the high importance regions at both low and high frequency regions. However, their overlap in the mid-frequency region could have contributed to a high number of misclassifications in the frequency limited case.

\vspace{-0.4 cm}
\subsection{Voiced vs Unvoiced Classification}
CAMs were utilized to investigate the region of importance in the spectrograms of voiced speech in distinguishing them from unvoiced speech. The results are shown in Fig. \ref{fig:vowels_CAM_unvoiced}.

The CAMs revealed that the region of importance for unvoiced speech predominantly lies below 700 Hz . Notably, this frequency range corresponds to the general location of fundamental frequency, that are typically absent in non-voiced speech. Though this  location is contrary to expectation as discussed in Section \ref{sec:format}, the model’s choice to focus on this disjoint spectral region indicates its discriminative power in accurately distinguishing between voiced and non-voiced speech. This observation is further supported by the excellent performance reflected in the confusion matrix as seen Fig. \ref {fig:vowels_CAM_4k} (f), where the model achieved an accuracy exceeding 98\%. Although there were some misclassifications observed between voiced and non-voiced speech samples, they were minimal.

\vspace{-0.4 cm}
\subsubsection{Discussion}
Observing all the CAMs, when all frequencies are available to the network to make decisions, the high vowels /\textipa{i}/ and /\textipa{u}/ used both the formant region ($<$ 4000 Hz) and spectral shape characteristics to make decisions. The low-back /\textipa{A}/ and mid vowel /\textipa{\textrhookrevepsilon}/ use only the formant region, while /\textipa{\ae}/ used both the formant region ($<$ 4000 Hz) and spectral shape characteristics to make decisions. When the frequency range was limited to 4000 Hz, majority of the vowels focused on the first and second formant frequency regions, which is less than 1500 Hz for /\textipa{i}/, /\textipa{A}/ and /\textipa{\ae}/. However, /\textipa{u}/ focused on a narrower low frequency $<$ 1000 Hz, where this vowel's first and second formants fall. /\textipa{\textrhookrevepsilon}/ assigns high importance to a region around 2000 Hz, which corresponds to its third formant. Observing the CAMs in the voiced vs unvoiced detection, it is clear that the region of importance corresponds to where the first four formants of voiced speech would lie in.

Overall, this analysis revealed that the neural network which was not pretrained on spectrograms, but only fine-tuned on them are focusing on formants in most cases to make their decisions. This is similar to what linguists would do. However, high frequencies are also considered in the decision making in some cases, which may not be needed depending on the task at hand. These high frequencies maybe a result of noise in the database. Based on this observation, there is scope to inform deep learning models on the region of interest to consider based on linguistics knowledge. 

\vspace{-0.4 cm}
\section{Conclusion}
In conclusion, this study explored the interpretation of spectrograms in machine learning for speech classification. The prediction results demonstrated higher accuracy in vowel classification models trained using ResNet-101. Insights were gained into what these networks "see" in spectrograms as classification cues.  However, CAM alone fell short in explaining misclassifications for certain vowels, highlighting challenges in capturing higher-level semantic concepts and abstract reasoning. The study also revealed unique activation characteristics in neural networks when distinguishing between voiced and unvoiced speech, focusing on formant regions. CAM served as an initial tool for interpretability, revealing regions contributing significantly to predictions. Future work should explore techniques like filter visualization, gradient-based visualization, activation maximization, occlusion analysis to improve understanding of deep learning models in speech classification. These techniques hold potential for unraveling network complexities, addressing higher-level semantic concept capture, and advancing interpretability and performance in speech classification tasks.
\label{sec:conclusion}

\bibliographystyle{splncs_srt}
\bibliography{author}

\end{document}